\documentclass[12pt]{article}

\usepackage[super]{cite}
\usepackage{graphicx}

\usepackage{amssymb}
\usepackage{color}   
\usepackage{hyperref}  
\hypersetup{colorlinks,urlcolor=blue,linkcolor=blue,citecolor=blue,filecolor=blue,urlcolor=blue}

\usepackage[latin1]{inputenc}

\newcommand{\newc}{\newcommand}
\newc{\beq}{\begin{equation}}
\newc{\eeq}{\end{equation}}
\newc{\bea}{\begin{array}}
\newc{\eea}{\end{array}}
\newcommand{\ben}{\begin{eqnarray}}
\newcommand{\een}{\end{eqnarray}}
\newc{\ra}{\rightarrow}
\newc{\bfx}{{\bf x}}
\newc{\bfV}{{\bf V}}
\newc{\cO}{{\cal O}}
\newc{\bfv}{{\bf v}}
\newc{\bfu}{{\bf u}}
\newc{\bfp}{{\bf p}}
\newc{\ve}{{\varepsilon}}
\newc{\Psibar}{\overline\Psi}
\newc{\w}{{\bf w}}
\newc{\E}{{\mathbf{E}}}
\newc{\EE}{{\mathcal E}}
\newc{\bfn}{{\mathbf\nabla}}
\newc{\la}{{\cal L}}
\newc{\tla}{{\tilde{\cal L}}}
\newc{\bp}{{\bf p}}
\newc{\ho}{\hookrightarrow }
\newc{\bP}{{\bf P}}
\newc{\pd}{{\partial}}
\newc{\piv}{{\partial_4}}
\newc{\pv}{{\partial_5}}
\newc{\bJ}{{\bf J}}
\newc{\bze}{{\mathbf 0}}
\newc{\bK}{{\bf K}}
\newc{\tphi}{{\tilde\phi}}
\newc{\tF}{{\tilde F}}
\newc{\tD}{{\tilde D}}
\newc{\tJ}{{\tilde J}}
\newc{\tj}{{\tilde j}}
\newc{\bD}{{\bf D}}
\newc{\tvphi}{{\tilde\varphi}}
\newc{\trho}{{\tilde\rho}}
\newc{\ttheta}{{\tilde\theta}}
\newc{\tpsi}{{\tilde\psi}}
\newc{\tu}{{\tilde u}}
\newc{\cD}{{\cal D}}
\newc{\tPhi}{{\tilde\Phi}}
\newc{\tPsi}{{\tilde\Psi}}
\newc{\tA}{{\tilde A}}
\newc{\talpha}{{\tilde\alpha}}
\newc{\tbeta}{{\tilde\beta}}
\newc{\bA}{{\mathbf A}}
\newc{\bB}{{\bf B}}
\newc{\br}{{\bf r}}
\newc{\sig}{{\mathbf\sigma}}
\newc{\eg}{{\rm e.g.\ }}
\newc{\ie}{{\rm i.e.\ }}
\newcommand{\bey}{\begin{eqnarray}}
\newcommand{\pslash}{\not{\hbox{\kern-2.3pt $p$}}}
\newcommand{\pdslash}{\not{\hbox{\kern-2pt $\partial$}}}
\newcommand{\eey}{\end{eqnarray}}

\begin{document}

\begin{titlepage}
\vskip 2cm
\begin{center}
{\Large  Clifford algebras, algebraic spinors, quantum information and applications
  \let\thefootnote\relax\footnotetext{$^{*}${\tt matrindade@uneb.br}}   }
 \vskip 10pt
{  Marco A. S. Trindade$^{1,*}$, Sergio Floquet$^{2}$ and J. David M. Vianna$^{3,4}$  \\}
\vskip 5pt
{\sl $^1$Departamento de Ciências Exatas e da Terra, Universidade do Estado da Bahia,
Colegiado de Física, Bahia, Brazil\\
$^2$Colegiado de Engenharia Civil, Universidade Federal do Vale do S\~ao Francisco, Brazil \\
$^3$Instituto de Física, Universidade Federal da Bahia, Brazil \\
$^4$Instituto de Física, Universidade de Brasília, Brazil}
\vskip 2pt

\end{center}

\begin{abstract}

We give an algebraic formulation based on Clifford algebras and algebraic spinors for quantum information. In this context, logic gates and concepts such as chirality, charge conjugation, parity and time reversal are introduced and explored in connection with states of qubits. Supersymmetry and M-superalgebra are also analysed with our formalism. Specifically we use extensively the algebras $Cl_{3,0}$ and $Cl_{1,3}$ as well as tensor products of Clifford algebras.
\end{abstract}

\bigskip



\vskip 3pt

\end{titlepage}


\newpage

\setcounter{footnote}{0} \setcounter{page}{1} \setcounter{section}{0} %
\setcounter{subsection}{0} \setcounter{subsubsection}{0}

\section{Introduction}
In relativistic quantum mechanics \cite{Dirac, Greiner}, Clifford algebras naturally appear through Dirac matrices. Covariant bilinears, chirality, CPT symmetries are some of the mathematical objects that play a fundamental role in the theory, built in terms of the spinors and generators of Dirac algebra. The ubiquitous character of Clifford algebras suggests the possibility of using them as a link between quantum computation \cite{Nielsen, Preskill} and high energy physics.
In fact, recently Martinez \emph{et al} \cite{Martinez} performed an experimental demonstration of a simulation of a network gauge theory using a low-q-trapped quantum ion computer. A relationship between the particle-antiparticle creation mechanism and the entanglement of the system, measured through logarithmic negativity, was also observed. Moreover there are several works in which Clifford algebra techniques are employed for quantum computing \cite{Havel1, Havel2, Baylis, Vlasov, Cabrera, Alves,Josza,topological1,Wein}. \

Another field in which Clifford algebras and spinors play a key role is supersymmetry \cite{Traubenberg, Varad}. An approach through spacetime algebra was obtained in \cite{Lasenby}.  A supersymmetric generalization of qubits has been proposed by \cite{Borsten} \emph{pari passu} to supersymmetric generalizations of entanglement measures through the concept of superdeterminants. Super Bell and super GHZ states have been defined. Local Lie Group Extensions Associated with Classical Communication Local Operations (LOCC), $\left [SU (2) \right]^{n}$, and Stochastic Classical Communication Local Operations (SLOCC) were performed with the $\left [Osp (1 \mid 2) \right]^{n}$ and $\left [uOsp (1 \mid 2) \right]^ {n}$ supergroups, respectively. \

In the M theory \cite{Duff}, Clifford algebras and octonions have been employed \cite{Anast, Toppan}to derive an eleven-dimensional supersymmetric algebra called M-algebra, a non-trivial extension of the Poincar\'e superalgebra. In eleven dimensions, the supercharge $Q$ is a pseudo-Majorana supercharge corresponds to a spinor with 32 components \cite{Traubenberg}. Indeed, possible connections to M-theory are investigated by Borsten \cite{BH}

In this work, we use algebraic spinors as main toll for the description of quantum information and their applications. Firstly, in the  section \ref{sec2} we present an algebraic formulation for multipartite q-bit systems in terms of the algebraic spinors in $ Cl_ {1,3} $. In section \ref{sec3} we derive a method for obtaining quantum logic gates in an algebraic setting.  In section \ref{sec4} we presents chirality, charge conjugation, parity and time reversal operators in algebraic formulation. In section \ref{sec5} we show how the supersymmetry can be explored in the context of spacetime algebra \cite{Lasenby}. Section \ref{sec6} contains  an approach to obtaining superalgebra by exploring its connection to octonions, qubits and entanglement. Finally, in section \ref{sec7}, we have the conclusions and perspectives.


\section{Algebraic spinors \label{sec2}}

In this section we will review some mathematical facts concerning algebraic spinors \cite{Vaz} and we show how to use them to describe qubits. So let us consider $A$ an associative algebra.  An left ideal $I_{L}$ is a subspace such that $AI_{L} \subset I_{L}$ and analogously, a vector subspace $I_{R}$ is a right ideal if $I_{R}A \subset I_{R}$. An element $\varepsilon \in A$ is said idempotent if $\varepsilon^{2}=\varepsilon$ and $\varepsilon \neq 0$. An idempotent is primitive if there are no idempotents $\varepsilon_{1}$ and $\varepsilon_{2}$ such that $\varepsilon_{1}\varepsilon_{2}=0$ and $\varepsilon_{1}+\varepsilon_{2}=\varepsilon$. The minimal left ideals in the Clifford algebra $Cl_{p,q}$ can be written in the form $Cl_{p, q}\varepsilon$. Algebraic spinors are defined as elements of minimal left ideals. It is worth mentioning that an alternative way is to use  q-bits in terms of operatorial spinors\cite{Havel1, Havel2} .
In this paper we will make use of Clifford algebra $Cl_{3,0}$
\begin{eqnarray}
\sigma_{i}\sigma_{j}+\sigma_{j}\sigma_{i}=2\delta_{ij}, \ \ i,j=1,2,3,
\end{eqnarray}
where $\delta_{ij}=0$ if $i\neq j$ and $\delta_{ij}=1$, otherwise.
Therefore, $\sigma_{i}\sigma_{j}=-\sigma_{j}\sigma_{i}, i \neq j$ and $\sigma_{i}^{2}=1$. Any element $\Gamma \in Cl_{3,0}$ can be written as a linear combination of $\{1\}$ (scalar), $ \sigma_{1}, \sigma_{2}, \sigma_{3}\}$ (vectors), $\{\sigma_{1}\sigma_{2}, \sigma_{2}\sigma_{3}, \sigma_{1}\sigma_{3}\}$, (bivectors) and
$\sigma_{1}\sigma_{2}\sigma_{3}$ (trivector). We will also use the Clifford algebra $Cl_ {1,3}$ (or spacetime algebra), defined as
\begin{eqnarray}
\gamma_{\mu}\gamma_{\nu}+\gamma_{\nu}\gamma_{\mu}=2g_{\mu \nu},
\end{eqnarray}
 where $g_{\mu \nu}$ is the Minkowiski metric with signature $(+---)$. Therefore, $\gamma_{\mu}\gamma_{\nu}=-\gamma_{\nu}\gamma_{\mu}, \mu \neq \nu$ and $\gamma_{0}^{2}=1, \gamma_{1}^{2}=\gamma_{2}^{2}= \gamma_{3}^{2}=-1$. Any element $\Gamma \in Cl_{1,3}$ can be written as a linear combination of $\{1\}$ (scalar), $\{\gamma_{0}, \gamma_{1}, \gamma_{2}, \gamma_{3}\}$ (vectors), $\{\gamma_{0}\gamma_{1}, \gamma_{0}\gamma_{2}, \gamma_{0}\gamma_{3}, \gamma_{1}\gamma_{2}, \gamma_{1}\gamma_{3}, \gamma_{2}\gamma_{3}\}$, (bivectors),
$\{\iota \gamma_{0}, \iota \gamma_{1}, \iota \gamma_{2}, \iota \gamma_{3}\}$ (pseudovectors) and $\{\iota \equiv \gamma_{0} \gamma_{1} \gamma_{2} \gamma_{3}\}$ (pseudoscalar). \

With the $Cl_{1,3}$ algebra, the Dirac equation can be written as \cite{Hestenes}
\begin{eqnarray}
\nabla \psi \iota \gamma_{3}\gamma_{0}=m \psi \gamma_{0}
\end{eqnarray}
where $\nabla= \gamma^{\mu}\partial_{\mu}$. Due to this fact we will use this algebra as a starting point for future developments in relativistic quantum information theory. \


Now we will build the isomorphism $Cl_{3,0}\simeq Cl_{1,3}^{+}$, $\xi: Cl_{3,0} \rightarrow Cl_{1,3}^{+}$, given by $\zeta(\sigma_{i})=\gamma_{i}\gamma_{0}$. We have the primitive idempotents
\begin{eqnarray}
\varepsilon & = & \frac{1}{2}(1+\sigma_{3}) \ \in Cl_{3,0} \nonumber \\
P & = & \frac{1}{2}(1+\gamma_{3}\gamma_{0}) \ \in Cl^{+}_{1,3} .
\end{eqnarray}
Thus we have the following correspondence:
\begin{eqnarray} \label{eq-5}
\left\lbrace \begin{array}{rccl}
 |0 \rangle & \leftrightarrow  & \sigma_{3}\varepsilon  &          \leftrightarrow    \gamma_{3}\gamma_{0}P,  \\
i|0 \rangle &  \leftrightarrow & \sigma_{1}\sigma_{2} \varepsilon  &   \leftrightarrow   \iota\gamma_{3} \gamma_{0}   P,
 \end{array}\right.  \hspace{1.5cm} \left\lbrace \begin{array}{rccl}
  |1 \rangle  & \leftrightarrow &  \sigma_{1}\varepsilon   &  \leftrightarrow   \gamma_{1}\gamma_{0}P, \\
  i|1 \rangle & \leftrightarrow & \sigma_{2}\sigma_{3}  \varepsilon  &  \leftrightarrow   \iota \gamma_{1} \gamma_{0}  P .
 \end{array} \right.
\end{eqnarray}


Therefore we can write the qubit as an algebraic spinor in $Cl_{3,0}$:
\begin{eqnarray} \label{estadoB}
\Psi &=&(\alpha_{1}\sigma_{3}+\alpha_{2}I \sigma_{3}+\alpha_{3}\sigma_{1}+\alpha_{4} I\sigma_{1})\varepsilon \nonumber \\
&=&(\alpha_{1}\sigma_{3}+\alpha_{2}\sigma_{1}\sigma_{2}+\alpha_{3}\sigma_{1}+\alpha_{4}\sigma_{2}\sigma_{3})\frac{1}{2}(1+\sigma_{3}), \label{eq6}
\end{eqnarray}
where $I=\sigma_{1}\sigma_{2}\sigma_{3}$ and $\alpha_{i} \in \mathbb{R}, \ i=1,2,3,4$. The corresponding state in the Hilbert space of (\ref{eq6}) is given by
\begin{eqnarray}
| \Psi \rangle = (\alpha_{1}+i \alpha_{2})|0 \rangle + (\alpha_{3}+i \alpha_{4})|1 \rangle.
\end{eqnarray}

\
Following this prescription we will consider multipartite q-bits as tensor products of algebraic spinors in $Cl_{3,0}$,i.e., $\left[Cl_{3,0}\right]^{\otimes n}$:
\begin{eqnarray}
\Psi &\in& \left[Cl_{3,0}\right]\varepsilon \otimes \left[Cl_{3,0}\right]\varepsilon \cdots \left[Cl_{3,0}\right]\varepsilon \nonumber \\
&=& \left[Cl_{3,0} \otimes Cl_{3,0} \otimes \cdots \otimes Cl_{3,0} \right][\varepsilon \otimes \varepsilon \otimes \cdots \otimes \varepsilon] \equiv  \left[Cl_{3,0}\right]^{\otimes n}\left[\varepsilon \right]^{\otimes n}.
\end{eqnarray}
For example, the bipartite state
\begin{eqnarray}
| \Psi \rangle & = & \left(\alpha_{1}  + i \alpha_{2} \right) |00 \rangle    +
\left(\alpha_{3}  + i \alpha_{4} \right) |01 \rangle  + \left(\alpha_{5}  + i \alpha_{6} \right) |10 \rangle
+ \left(\alpha_{7}  + i \alpha_{8} \right) |11 \rangle \nonumber \\
\end{eqnarray}
is associated to
\begin{eqnarray} \label{bipartite}
\Psi &=&  \left[\alpha_{1} (\sigma_{3})\otimes (\sigma_{3})
+ \alpha_{2}  (\sigma_{3} )\otimes (\sigma_{1}\sigma_{2}) \right. \nonumber \\
& & + \alpha_{3} (\sigma_{3})\otimes (\sigma_{1})
+ \alpha_{4} (\sigma_{3})\otimes (\sigma_{2}\sigma_{3}) \nonumber \\
& & + \alpha_{5} (\sigma_{1})\otimes (\sigma_{3})
+ \alpha_{6} (\sigma_{1})  \otimes  (\sigma_{1}\sigma_{2}) \nonumber \\
& & \left. + \alpha_{7} (\sigma_{1})\otimes (\sigma_{1})
+ \alpha_{8} (\sigma_{1})\otimes (\sigma_{2}\sigma_{3}) \right] \varepsilon^{\otimes 2}.
\end{eqnarray}
with $\alpha_{i} \in \mathbb{R}, \ i=1,2,...,8$. We define an entangled bipartite state as a state that cannot be written in the form $\Psi= \Psi_{1} \otimes \Psi_{2}$, otherwise, the state is said to be separable.


\section{Algebraic quantum logic gates \label{sec3}}
We can obtain quantum logic gates based on
algebraic elements. For this purpose we will use the algebra $Cl_{3,0}$. Our construction
makes use of the fact that quantum logic gates are elements
of the Lie group  $ U (n) $ and that there is a
isomorphism between subalgebras of tensor product of algebras $Cl_{3,0}$ and the Lie algebra of the group $U(2^{n})$, denoted by $u(2^{n})$.
The Lie algebra $u(n)$ consists of the space of anti-hermitian complex matrices over $\mathbb{R}$.
An arbitrary element of $u (n)$ has the form \cite{Lie4}
\begin{equation}
M=\sum_{i,j=1}^{n} \alpha _{ij}E_{ij},
\end{equation}%
where $E_{ij}$ are matrices of dimension $n\times n$ with the element
$i,j$ equal to $1$, and all the other elements are zero. 
The elements $\alpha _{ij}$ are coefficients belonging to the field of complex numbers and must satisfy the relation
\begin{equation}
\alpha _{ij}=-\alpha _{ji}^{\ast }.
\end{equation}%
For the Lie algebra $u(n)$, we have the generators
\begin{eqnarray}
M_{ij} &=&i(E_{ij}+E_{ji}),\ \ \ if \ \ i>j, \nonumber  \\
M_{ij} &=&E_{ij}-E_{ji},\ \ \ \ \ \ \ if\ \ i<j, \nonumber \\
M_{ii} &=&iE_{ii}.
\end{eqnarray}%

Note that the elements $M_{ij}$ and $M_{ii}$ are anti-hermitian and
linearly independent. For example, in the algebra $u(2)$ we have
\begin{eqnarray}
M_{11}=\left(
\begin{array}{cc}
i & 0 \\
0 & 0%
\end{array}%
\right) ,\ \ \ \ \ \ \ \ \ \ \ M_{12}=\left(
\begin{array}{cc}
0 & 1 \\
-1 & 0%
\end{array}%
\right),
\end{eqnarray}%
\begin{eqnarray}
M_{21}=\left(
\begin{array}{cc}
0 & i \\
i & 0%
\end{array}%
\right) ,\ \ \ \ \ \ \ \ \ \ \ M_{22}=\left(
\begin{array}{cc}
0 & 0 \\
0 & i%
\end{array}%
\right).
\end{eqnarray}

It is possible to perform a change of basis on the algebra $u(2)$, building a new base in terms of multiples of the Pauli matrices $\left\lbrace iX,  iY, iZ, i1 \right\rbrace $:
\begin{eqnarray}
iX=\left(
\begin{array}{cc}
0 & i \\
i & 0%
\end{array}%
\right) ,\ \ \ \ \ \ \ \ \ iY=\left(
\begin{array}{cc}
0 & 1 \\
-1 & 0%
\end{array}%
\right),
\end{eqnarray}%
\begin{eqnarray}
iZ=\left(
\begin{array}{cc}
i & 0 \\
0 & -i%
\end{array}%
\right) ,\ \ \ \ \ \ \ \ \ i1=\left(
\begin{array}{cc}
i & 0 \\
0 & i%
\end{array}%
\right).
\end{eqnarray}%
Note that if we exclude the element $i1$, we have the Lie algebra $su(2)$
. In terms of the algebra $Cl_{3,0}$, whose general element (multivector) is
given by
\begin{equation}
\sigma = \alpha^{0}1+\alpha^{1}\sigma _{1}+\alpha^{2}\sigma _{2}+\alpha^{3}\sigma_{3}+\alpha^{12}\sigma_{1}\sigma_{2}+\alpha^{23}\sigma_{2}\sigma_{3}+\alpha^{31}\sigma_{3}\sigma_{1}+\alpha^{123}\sigma_{1}\sigma_{2}\sigma_{3},
\label{multivetorgama}
\end{equation}%
we can build generators for $U(2)$ through the algebra of
bivectors, added to the trivetor $ I = \sigma_{1}\sigma_{2}\sigma_{3}$, i.e.
\begin{equation}
\left\{ \sigma_{1}\sigma_{2},\sigma_{2}\sigma_{3},\sigma_{3}\sigma
_{1},\sigma_{1}\sigma_{2}\sigma_{3}\right\}.
\end{equation}

These elements form an algebra
\begin{eqnarray}
\left[ \sigma_{1}\sigma_{2},\sigma_{2}\sigma_{3}\right] &=&-2\sigma
_{3}\sigma_{1},\  \ \  \ \ \left[ \sigma_{1}\sigma_{2},\sigma_{3}\sigma_{1}%
\right] \ =2\ \sigma_{2}\sigma_{3}\ ,\ \ \ \left[ \sigma_{1}\sigma
_{2},\sigma_{1}\sigma_{2}\sigma_{3}\right] =0, \nonumber \\
\left[ \sigma_{2}\sigma_{3},\sigma_{3}\sigma_{1}\right] &=&-2\sigma
_{1}\sigma_{2},\  \ \ \left[ \sigma_{2}\sigma_{3},\sigma_{1}\sigma
_{2}\sigma_{3}\right] =0,\ \ \ \ \ \ \ \ \ \ \ \left[ \sigma_{3}\sigma
_{1},\sigma_{1}\sigma_{2}\sigma_{3}\right] =0. \nonumber \\
\end{eqnarray}

We have four generators and this algebra is isomorphic to $U(2)$ Lie algebra. Excluding the trivetor $\sigma_{1}\sigma _{2}\sigma_{3}$, which is the only constituent of the center of algebra, we have an
isomorphism with $SU(2)$ Lie algebra. For $u(4)$, we have $16$
generators; these generators can be built as:
\begin{eqnarray}
&&\sigma_{1}\sigma_{2}\otimes \sigma_{1},\ \ \ \ \ \ \  \ \ \ \sigma
_{1}\sigma_{2}\otimes \sigma_{2},\ \ \ \ \ \ \ \  \ \ \ \sigma
_{1}\sigma_{2}\otimes \sigma_{3},\ \ \ \ \ \ \ \  \ \ \sigma_{1}\sigma
_{2}\otimes 1, \nonumber \\
&&\sigma_{2}\sigma_{3}\otimes \sigma_{1},\ \ \ \ \  \ \ \ \ \ \sigma
_{2}\sigma_{3}\otimes \sigma_{2},\ \ \ \ \ \ \ \  \ \ \ \sigma
_{2}\sigma_{3}\otimes \sigma_{3},\ \ \ \ \ \ \ \  \ \ \sigma_{2}\sigma
_{3}\otimes 1,\  \nonumber \\
&&\sigma_{3}\sigma_{1}\otimes \sigma_{1},\ \ \ \  \ \ \ \ \ \ \sigma
_{3}\sigma_{1}\otimes \sigma_{2},\ \ \ \ \ \ \ \  \ \ \ \sigma
_{3}\gamma _{1}\otimes \sigma_{3},\ \ \ \ \ \ \ \ \ \ \sigma_{3}\sigma
_{1}\otimes 1, \nonumber \\
&&\sigma_{1}\sigma_{2}\sigma_{3}\otimes \sigma_{1},\ \  \ \ \ \ \sigma
_{1}\sigma_{2}\sigma_{3}\otimes \sigma_{2},\ \ \  \ \ \ \ \sigma
_{1}\sigma_{2}\sigma_{3}\otimes \sigma_{3},\ \ \  \ \ \ \sigma
_{1}\sigma_{2}\sigma_{3}\otimes 1.
\end{eqnarray}%

We define the reversal operation of a multivector observing that the elements above are linearly independent and the tensor product between bivectors and vectors for a representation
in terms of Pauli's matrices results in anti-hermitian matrices; a same result
 occurs for the tensor product between a trivetor (or bivector) and the
identity. If we are interested in $su(4)$ algebra, we should exclude
the element $1\otimes \sigma_{1}\sigma_{2}\sigma_{3}$. For the
$u(8)$ algebra, we have $ 64 $ generators obtained by taking tensor products
between bivectors (or trivectors) and vectors (identity), with the condition
that the bivectors (trivectors) appear an odd number of
times. This can be better understood if we refer to the reversal operation in a Clifford algebra, in analogy with the matrix Hermitian operators.
We define the reversion operation of a multivetor $%
\Gamma $ (a $p$-vetor) as
\begin{equation}
\widetilde{\Gamma}_{\left[ p\right] }=\left( -1\right)
^{p(p-1)/2}\Gamma _{\left[ p\right] }.
\end{equation}%
For a scalar, a vector, a bivetor and a trivector, we have, respectively%
\begin{equation}
\widetilde{\Gamma}_{\left[ 0\right] }=\Gamma _{\left[ 0\right] },\ \ \
\widetilde{\Gamma}_{\left[ 1\right] }=\Gamma _{\left[ 1\right] },\ \
\widetilde{\Gamma}_{\left[ 2\right] }=-\Gamma _{\left[ 2\right] },\ \
\ \ \ \widetilde{\Gamma}_{\left[ 3\right] }=-\Gamma _{\left[ 3\right] }.
\end{equation}%

The analogy can be seen if we consider a matrix representation for the multivector  $\Gamma $, given by (\ref{multivetorgama}%
), in terms of Pauli matrices,%
\begin{eqnarray}
\Gamma =\left(
\begin{array}{cc}
(\alpha ^{0}+\alpha ^{3})+i(\alpha ^{12}+\alpha ^{123}) & (\alpha
^{1} + \alpha ^{31})+i(\alpha ^{23}-\alpha ^{2})  \\
(\alpha ^{1}-\alpha ^{31})+i(\alpha ^{23}+\alpha ^{2}) & (\alpha ^{0}-\alpha
^{3})+i(\alpha ^{123}-\alpha ^{12})%
\end{array}%
\right) =\left(
\begin{array}{cc}
z_{1} & z_{2}  \\
z_{3} & z_{4}%
\end{array}%
\right) . \nonumber \\
\end{eqnarray}%
Performing the reversion operation, we have
\begin{eqnarray}
\widetilde{\Gamma}=\left(
\begin{array}{cc}
\widetilde{{z}_{1}} & \widetilde{{z}_{3}} \nonumber \\
\widetilde{{z}_{2}} & \widetilde{{z}_{4}}%
\end{array}%
\right) ,
\end{eqnarray}%
that corresponds to conjugate transpose of $\Gamma $. Taking the tensor product of bivectors, we obtain
\begin{eqnarray}
\widetilde{\Gamma }_{\left[ 2\right] }\otimes \widetilde{\Gamma }_{%
\left[ 2\right] }=(-\Gamma _{\left[ 2\right] })\otimes (-\Gamma _{\left[ 2%
\right] })=\Gamma _{\left[ 2\right] }\otimes \Gamma _{\left[ 2\right] },
\end{eqnarray}%
i.e. we have a Hermitian element. On the other hand, carrying out the
tensor product of a bivector with a vector, we have%
\begin{eqnarray}
\widetilde{\Gamma }_{\left[ 2\right] }\otimes \widetilde{\Gamma }%
_{\left[ 1\right] }=(-\Gamma _{\left[ 2\right] })\otimes \Gamma _{\left[ 1%
\right] }=-\Gamma _{\left[ 2\right] }\otimes \Gamma _{\left[ 1\right] },
\end{eqnarray}%
that result is anti-hermitian element; a similar analysis is valid for trivector. In the general case of algebra $u(2^{n})$, we have%
\begin{eqnarray}
\sigma_{1}\sigma_{2} & \otimes ...\otimes & \ \sigma_{1},\ \ \ \ \ ...\ \ \ \ \sigma_{1}\sigma_{2} \otimes ...\otimes \ 1 , \nonumber \\
\sigma_{2}\sigma_{3} & \otimes ...\otimes & \ \sigma_{1}, \ \ \ \ \ ...\ \ \ \ \sigma_{2}\sigma_{3}\otimes ...\otimes \ 1,  \nonumber \\
\sigma_{3}\sigma_{1} & \otimes ...\otimes & \ \sigma_{1},\ \ \ \ \ ... \ \ \ \ \sigma_{3}\sigma_{1}\otimes ...\otimes \ 1, \nonumber \\
\sigma_{1}\sigma_{2}\sigma_{3} & \otimes ... \otimes & \ \sigma_{1}, \ \ \ \ \ ... \ \ \ \ \sigma_{1}\sigma_{2}\sigma_{3}\otimes ...\otimes 1 ,
\end{eqnarray}%
with bivectors and trivectors appearing an odd number of times,
in order to ensure that the tensor product results in an anti-hermitian element.
For example, in the case of algebra $u(2^{3})=u(8)$, the terms in which
the first factor of the tensor product is $\sigma_{1}\sigma_{2}$, and the
last factor is $\sigma_{1}$, following an analogous procedure to that of $u(4)$, we get
\begin{eqnarray}
\sigma_{1}\sigma_{2}\otimes \sigma_{1}\otimes \ \sigma_{1}, \\
\sigma_{1}\sigma_{2}\otimes \sigma_{2}\otimes \ \sigma_{1}, \\
\sigma_{1}\sigma_{2}\otimes \sigma_{3}\otimes \ \sigma_{1}, \\
\sigma_{1}\sigma_{2}\otimes 1\otimes \ \sigma_{1}.
\end{eqnarray}%

The exponential of a multivector $\Gamma$ is defined as
\begin{equation}
\exp \Gamma =\sum_{n=0}^{\infty}\frac{\Gamma ^{n}}{n!}.
\end{equation}%
Then, for a multivector $\Gamma $
\begin{eqnarray}
\Gamma  &=&\alpha _{12}^{...1}\sigma_{1}\sigma_{2}\otimes ...\otimes \ \sigma_{1}+...
+\alpha _{12}^{...0}\sigma_{1}\sigma_{2}\otimes ...\otimes \ 1 \\
&& + \alpha _{23}^{...1}\sigma_{2}\sigma_{3}\otimes ...\otimes \ \sigma_{1}+...
+\alpha _{23}^{...0}\sigma_{2}\sigma_{3}\otimes ...\otimes \ 1 \nonumber \\
&& + \alpha _{31}^{...1}\sigma_{3}\sigma_{1}\otimes ...\otimes \ \sigma_{1}+...
+\alpha _{31}^{...0}\sigma_{3}\sigma_{1}\otimes ...\otimes \ 1 \nonumber \\
&& + \alpha _{0}^{...31}1\otimes ...\otimes \sigma_{3}\sigma_{1} +... +\alpha_{0}^{...123}1\otimes ...\otimes \sigma_{1}\sigma_{2}\sigma_{3},  \nonumber
\end{eqnarray}%
a general element of $U(2^{n})$ is given by
\begin{eqnarray}
U &=&\exp \left(\alpha _{12}^{...1}\sigma_{1}\sigma_{2}\otimes ...\otimes \
\sigma_{1}+...+\alpha _{12}^{...0}\sigma_{1}\sigma_{2}\otimes ...\otimes
\ 1 \right.  \nonumber \\
&&+\alpha _{23}^{...1}\sigma_{2}\sigma_{3}\otimes ...\otimes \ \sigma
_{1}+...+\alpha _{23}^{...0}\sigma_{2}\sigma_{3}\otimes ...\otimes \ 1
\nonumber \\
&&+\alpha _{31}^{...1}\sigma_{3}\sigma_{1}\otimes ...\otimes \ \sigma
_{1}+...+\alpha _{31}^{...0}\sigma_{3}\sigma_{1}\otimes ...\otimes \ 1
\nonumber \\
&& \left. +\alpha _{0}^{...31}1\otimes ...\otimes \sigma_{3}\sigma
_{1}+...+ \alpha_{0}^{...123}1\otimes ...\otimes \sigma_{1}\sigma_{2}\sigma_{3} \right).
\end{eqnarray}%
where the lower indices refer to the first factor of the
tensor product, assigning the index $0$ for the unit,
and the upper indices indicate the sequence in which the multivectors appear at
from the second factor. The $\alpha_{i}^{j}$ coefficients are real.

With this mathematical development we show how it is possible to build logic gates
within the algebraic formulation through a $U(2^{n})$ Lie algebra structure contained within
of the algebra product $ Cl_{3,0} \otimes Cl_{3,0} \otimes ... \otimes Cl_{3,0} $.
These logic gates can be expressed through the algebra generators. In fact, when these logic gates
operationalize over the states, the resultant state can be obtained easily since we have
relations between the generators and the states are also expressed in terms of these generators.
Consider initially, the quantum logic gate
$NOT$ denoted by the Pauli matrix  $X$, whose effect on a
qubit is given by
\begin{eqnarray}
X\left(
\begin{array}{c}
\alpha  \nonumber \\
\beta
\end{array}%
\right) =\left(
\begin{array}{c}
\beta  \nonumber \\
\alpha
\end{array}%
\right) .
\end{eqnarray}%
Since $Cl_{3,0}$ is defined on field of reals, we should
write the state of qubit as%
\begin{eqnarray}
\psi =(\alpha _{1}+\alpha _{2}I)\sigma_{3}\varepsilon _{1}+(\beta
_{1}+\beta _{2}I)\sigma_{1}\varepsilon _{1}.
\end{eqnarray}%
Then
the element associated to gate $X$ is given by $\sigma
_{1}$, since%
\begin{eqnarray}
\sigma_{1} \left[ (\alpha _{1}+\alpha _{2}I)\sigma_{3}\varepsilon _{1}+(\beta
_{1}+\beta _{2}I)\sigma_{13}\varepsilon _{1} \right] 
&=&(\beta _{1}+\beta _{2}I)\sigma_{3}\varepsilon _{1}+(\alpha _{1}+\alpha
_{2}I)\sigma_{1}\varepsilon _{1}. \nonumber \\
\end{eqnarray}%

The element $\sigma_{1}$ can be obtained through of product%
\begin{eqnarray}
U_{x} &=&\exp (\alpha \sigma_{1}\sigma_{2}\sigma_{3})\exp (\theta \sigma
_{2}\sigma_{3}) \nonumber \\
&=&\exp (\alpha \sigma_{1}\sigma_{2}\sigma_{3}+\theta \sigma_{2}\sigma
_{3}),
\end{eqnarray}%
where $\alpha =-\pi /2,$ $\theta =\pi /2$ and the second equality occurs because the trivetor $\sigma_{1}\sigma_{2}\sigma _{3}$ commutes with the bivector
 $\sigma_{2}\sigma_{3}$. To better understand this result,
we will analyze each one of the above factors. As $U_{x_{1}}=\exp (\alpha \sigma
_{1}\sigma_{2}\sigma_{3})$, we have%
\begin{eqnarray}
U_{x_{1}} &=&\exp (\alpha \sigma_{1}\sigma_{2}\sigma_{3}) \nonumber \\
&=&1+\alpha \sigma_{1}\sigma_{2}\sigma_{3}+\frac{\alpha ^{2}}{2!}\left(
\sigma_{1}\sigma_{2}\sigma_{3}\right) ^{2}+\frac{\alpha ^{3}}{3!}\left(
\sigma_{1}\sigma_{2}\sigma_{3}\right) ^{3}+... \nonumber \\
&=&1+\alpha \sigma_{1}\sigma_{2}\sigma_{3}-\frac{\alpha ^{2}}{2!}1-\frac{%
\alpha ^{3}}{3!}\sigma_{1}\sigma_{2}\sigma_{3}+\frac{\alpha ^{4}}{4!}1+%
\frac{\alpha ^{5}}{5!}\sigma_{1}\sigma_{2}\sigma_{3} \nonumber \\
&=&(1-\frac{\alpha ^{2}}{2!}+\frac{\alpha ^{4}}{4!}-...)+(\alpha -\frac{%
\alpha ^{3}}{3!}+\frac{\alpha ^{5}}{5!}-...)\sigma_{1}\sigma_{2}\sigma_{3}
\nonumber \\
&=&\cos (\alpha )1+\sin(\alpha )\sigma_{1}\sigma_{2}\sigma_{3}.
\end{eqnarray}%
For $U_{x_{2}}=\exp (\theta \sigma_{2}\sigma_{3}),$ we obtain
\begin{eqnarray}
U_{x_{2}} &=&\exp (\theta \sigma_{2}\sigma_{3})= \nonumber \\
&=&1+\theta \sigma_{2}\sigma_{3}+\frac{\theta ^{2}}{2!}\left( \sigma
_{2}\sigma_{3}\right) ^{2}+\frac{\theta ^{3}}{3!}\left( \sigma_{2}\sigma
_{3}\right) ^{3}+... \nonumber \\
&=&1+\theta \sigma_{2}\sigma_{3}-\frac{\theta ^{2}}{2!}1-\frac{\theta ^{3}%
}{3!}\sigma_{2}\sigma_{3}+\frac{\theta ^{4}}{4!}1+\frac{\theta ^{5}}{5!}%
\sigma_{2}\sigma_{3} \nonumber \\
&=&(1-\frac{\theta ^{2}}{2!}+\frac{\theta ^{4}}{4!}-...)+(\theta -\frac{%
\theta ^{3}}{3!}+\frac{\theta ^{5}}{5!}-...)\sigma_{2}\sigma_{3} \nonumber \\
&=&\cos (\theta )1+\sin(\theta )\sigma_{2}\sigma_{3}.
\end{eqnarray}%
Then, with  $\alpha =-\pi /2$ and $\theta =\pi /2$ we get
\begin{eqnarray}
U &=&U_{x_{1}}U_{x_{2}} \nonumber \\
&=&\sin(-\pi /2)\ \sigma_{1}\sigma_{2}\sigma_{3}\sin(\pi /2)\sigma_{2}\sigma_{3} \nonumber \\
&=&\sigma_{1}.
\end{eqnarray}%
Another logical gate of interest is the gate
$CNOT$, which performs the transformation
\begin{eqnarray}
\left\vert 00\right\rangle  &\rightarrow &\left\vert 00\right\rangle , \nonumber \\
\left\vert 01\right\rangle  &\rightarrow &\left\vert 01\right\rangle , \nonumber \\
\left\vert 10\right\rangle  &\rightarrow &\left\vert 11\right\rangle , \nonumber \\
\left\vert 11\right\rangle  &\rightarrow &\left\vert 10\right\rangle .
\end{eqnarray}%
To determine the algebraic element in this case, consider
\begin{eqnarray}
U &=&\exp [\alpha (\sigma_{1}\sigma_{2}\sigma_{3}\otimes 1)] \nonumber \\
&&\times \exp [\frac{\theta }{2}(\sigma_{1}\sigma_{2}\sigma_{3}\otimes
1+\sigma_{1}\sigma_{2}\otimes 1+\sigma_{1}\sigma_{2}\sigma_{3}\otimes
\sigma_{1}-\sigma_{1}\sigma_{2}\otimes \sigma_{1})].
\end{eqnarray}%
Since
\begin{eqnarray}
\left( \sigma_{1}\sigma_{2}\sigma_{3}\otimes 1\right) ^{2} &=&-1\otimes 1
\end{eqnarray}
and
\begin{eqnarray}
\hspace{-0.4cm} \left[ \frac{1}{2}(\sigma_{1}\sigma_{2}\sigma_{3}\otimes 1+\sigma
_{1}\sigma_{2}\otimes 1+\sigma_{1}\sigma_{2}\sigma_{3}\otimes \sigma_{1}-
\sigma_{1}\sigma_{2}\otimes \sigma_{1})\right] ^{2} &=&-1\otimes 1,
\end{eqnarray}%
we can rewrite $U$ by performing a procedure analogous to that of the logic gate $X$, i.e.
\begin{eqnarray}
\hspace{-0.4cm} U &=& \left[ \cos \alpha (1\otimes 1)+ \sin\alpha \lbrack (\sigma_{1}\sigma_{2}\sigma_{3}\otimes 1) \right] \left[\cos \theta (1\otimes 1) \right. \nonumber \\
\hspace{-0.4cm} &&  +\frac{1}{2} \left. \sin\theta  (\sigma_{1}\sigma_{2}\sigma_{3}\otimes
1+\sigma_{1}\sigma_{2}\otimes 1+\sigma_{1}\sigma_{2}\sigma_{3}\otimes
\sigma_{1}-\sigma_{1}\sigma_{2}\otimes \sigma_{1}) \right].
\end{eqnarray}%
And, if we set $\alpha =-\pi /2$ and $\theta =\pi /2$, we have
\begin{eqnarray}
U &=&-(\sigma_{1}\sigma_{2}\sigma_{3}\otimes 1) \left[ \frac{1}{2}(\sigma
_{1}\sigma_{2}\sigma_{3}\otimes 1+\sigma_{1}\sigma_{2}\otimes 1+\sigma
_{1}\sigma_{2}\sigma_{3}\otimes \sigma_{1}-\sigma_{1}\sigma_{2}\otimes
\sigma_{1}) \right] \nonumber \\
&=&\frac{1}{2}(1\otimes 1+\sigma_{3}\otimes 1+1\otimes \sigma_{1}-\sigma
_{3}\otimes \sigma_{1}) \nonumber \\
&=&\frac{1}{2} \left[ (1+\sigma_{3})\otimes 1+(1-\sigma_{3})\otimes \sigma_{1} \right]
\nonumber \\
&=&\frac{1}{2}(1+\sigma_{3})\otimes 1+\frac{1}{2}(1-\sigma_{3})\otimes
\sigma_{1}.
\end{eqnarray}%
This $U$ is the algebraic element correspondent of the gate $CNOT$. In fact,%
\begin{eqnarray}
U(\sigma_{3}\otimes \sigma_{3})(\varepsilon _{1}\otimes \varepsilon _{1})
&=&(\sigma_{3}\otimes \sigma_{3})(\varepsilon _{1}\otimes \varepsilon
_{1}), \nonumber \\
U(\sigma_{3}\otimes \sigma_{1})(\varepsilon _{1}\otimes \varepsilon _{1})
&=&(\sigma_{3}\otimes \sigma_{1})(\varepsilon _{1}\otimes \varepsilon
_{1}), \nonumber \\
U(\sigma_{1}\otimes \sigma_{3})(\varepsilon _{1}\otimes \varepsilon _{1})
&=&(\sigma_{1}\otimes \sigma_{1})(\varepsilon _{1}\otimes \varepsilon
_{1}), \nonumber \\
U(\sigma_{1}\otimes \sigma_{1})(\varepsilon _{1}\otimes \varepsilon _{1})
&=&(\sigma_{1}\otimes \sigma_{3})(\varepsilon _{1}\otimes \varepsilon
_{1}).
\end{eqnarray}

An important aspect to note in our algebraic formulation is that we obtained two logic gates as
examples - the logic gate $(X)$ and the logic gate $CNOT$. The others
logic gates of a q-bit, $(Y)$ and $(Z)$, can be generated with an
identical procedure to that used for gate $ (X) $. In consequence, our results are that its possible to build
the logic gates of a q-bit and the logical gate $CNOT$ algebraically. Hence, as any other quantum logic gate can be built from these gates\cite{Nielsen}, it follows that our proposal for quantum algebraic computing can be interest to this research area.


\section{Chirality, charge conjugation, parity and time reversal operators \label{sec4}}

In this section we explore some concepts of the relativistic quantum mechanics by using  $Cl_{1,3}^{+}\simeq Cl_{3,0}$ 
and algebraic spinors. Similar results were obtained with Dirac algebra \cite{Raja}. Here however we explore the structure of Clifford algebra and algebraic spinors. For example, the chirality operator $\gamma_{5}=i\gamma_{0}\gamma_{1}\gamma_{2}\gamma_{3}$, that in terms of Weyl representation is given by
\begin{eqnarray}
\gamma _{5}=\left(
\begin{array}{cc}
1_{2 \times 2} & 0 \\
0 & -1_{2 \times 2}%
\end{array}%
\right),
\end{eqnarray}
in our formulation $\gamma_{5}$ can be written as
\begin{eqnarray}
\Gamma_{5}^{Cl_{1,3}^{+}\otimes Cl_{1,3}^{+}}=\gamma_{3}\gamma_{0} \otimes 1=\widetilde{\Gamma}_{5}^{Cl_{1,3}^{+}\otimes Cl_{1,3}^{+}}.
\end{eqnarray}
The spinors that are eigenvectors of $\Gamma_{5}$ with eigenvalues $1$ and $-1$ are called right-handed spinors $\Psi_{R}$, and left-handed spinors $\Psi_{L}$, respectively:
\begin{eqnarray}
\Gamma_{5}^{Cl_{1,3}^{+}\otimes Cl_{1,3}^{+}}\Psi_{R}^{Cl_{1,3}^{+}}&=&\Psi_{R}^{Cl_{1,3}^{+}}, \nonumber \\
\Gamma_{5}^{Cl_{1,3}^{+}\otimes Cl_{1,3}^{+}}\Psi_{L}^{Cl_{1,3}^{+}}&=& - \Psi_{L}^{Cl_{1,3}^{+}}.
\end{eqnarray}

We can write a general state of a qubit as an algebraic spinor in $Cl_{1,3}^{+}\otimes Cl_{1,3}^{+} $ using the prescription of (\ref{eq-5}),
\begin{eqnarray} \label{estadoB}
\hspace{-0.4cm}\Psi^{Cl_{1,3}^{+}} &=&(\alpha_{1}\gamma_{3}\gamma_{0}+\alpha_{2}\iota\gamma_{3}\gamma_{0}+\alpha_{3}\gamma_{1}\gamma_{0}+\alpha_{4}\iota\gamma_{1}\gamma_{0})P \nonumber \\
\hspace{-0.4cm}&=&(\alpha_{1}\gamma_{3}\gamma_{0}+\alpha_{2}\gamma_{1}\gamma_{0}\gamma_{2}\gamma_{0}+\alpha_{3}\gamma_{1}\gamma_{0}+\alpha_{4}\gamma_{2}\gamma_{0}\gamma_{3}\gamma_{0})\frac{1}{2}(1+\gamma_{3}\gamma_{0}), \label{eq-51}
\end{eqnarray}
with $\alpha_{i} \in \mathbb{R}, \ i=1,2,3,4$. Eq. (\ref{eq-51}) is an element of the minimal left ideal $Cl_{1,3}^{+}P$ and corresponds to Hilbert state
\begin{eqnarray}
| \Psi^{Cl_{1,3}^{+}} \rangle = (\alpha_{1}+i \alpha_{2})|0 \rangle + (\alpha_{3}+i \alpha_{4})|1 \rangle.
\end{eqnarray}

In the multipartite systems we need to build the product algebra $\left[Cl_{1,3}^{+}\right]^{\otimes n}$.
We can write multipartite q-bits as tensor products of algebraic spinors in $Cl_{1,3}^{+}$, i.e. as elements of minimal left ideals in $\left[Cl_{1,3}^{+}\right]^{\otimes n}$:
\begin{eqnarray}
\Psi^{\left[Cl_{1,3}^{+}\right]^{\otimes n}} &\in& \left[Cl_{1,3}^{+}\right]P \otimes \left[Cl_{1,3}^{+}\right]P \cdots \left[Cl_{1,3}^{+}\right]P \nonumber \\
&=& \left[Cl_{1,3}^{+} \otimes Cl_{1,3}^{+} \otimes \cdots \otimes Cl_{1,3}^{+} \right][P \otimes P \otimes \cdots \otimes P] \nonumber \\
&\equiv & \left[Cl_{1,3}^{+}\right]^{\otimes n}\left[P \right]^{\otimes n}.
\end{eqnarray}
so that a bipartite state
\begin{eqnarray}
| \Psi \rangle & = & \left(\alpha_{1}  + i \alpha_{2} \right) |00 \rangle    +
\left(\alpha_{3}  + i \alpha_{4} \right) |01 \rangle  + \left(\alpha_{5}  + i \alpha_{6} \right) |10 \rangle
+ \left(\alpha_{7}  + i \alpha_{8} \right) |11 \rangle \nonumber \\
\end{eqnarray}
corresponds to
\begin{eqnarray} \label{bipartite1}
\Psi^{Cl_{1,3}^{+} \otimes Cl_{1,3}^{+} } &=&  \left[\alpha_{1} (\gamma_{3}\gamma_{0} \otimes \gamma_{3}\gamma_{0})
+ \alpha_{2}  (\gamma_{3}\gamma_{0} \otimes \gamma_{1}\gamma_{0}\gamma_{2}\gamma_{0}) \right. \nonumber \\
& & + \alpha_{3} (\gamma_{3}\gamma_{0} \otimes \gamma_{1}\gamma_{0})
+ \alpha_{4} (\gamma_{3}\gamma_{0} \otimes \gamma_{2}\gamma_{0}\gamma_{3}\gamma_{0}) \nonumber \\
& & + \alpha_{5} (\gamma_{1}\gamma_{0} \otimes \gamma_{3}\gamma_{0})
+ \alpha_{6} (\gamma_{1}\gamma_{0}   \otimes   \gamma_{1}\gamma_{0}\gamma_{2}\gamma_{0}) \nonumber \\
& & \left. + \alpha_{7} (\gamma_{1}\gamma_{0} \otimes \gamma_{1}\gamma_{0})
+ \alpha_{8} (\gamma_{1}\gamma_{0} \otimes \gamma_{2}\gamma_{0}\gamma_{3}\gamma_{0}) \right] P^{\otimes 2}.
\end{eqnarray}
where $\alpha_{i} \in \mathbb{R}, \ i=1,2,...,8$.

The chiral projection operators are given by
\begin{eqnarray}
P^{R}&=&\frac{1}{2}\left(1+\Gamma_{5}^{Cl_{1,3}^{+}\otimes Cl_{1,3}^{+}} \right)=\frac{1}{2}\left(1 \otimes 1+ \gamma_{3}\gamma_{0}\otimes 1 \right) \nonumber \\
P^{L}&=&\frac{1}{2}\left(1-\Gamma_{5}^{Cl_{1,3}^{+}\otimes Cl_{1,3}^{+}} \right)=\frac{1}{2}\left(1 \otimes 1- \gamma_{3}\gamma_{0}\otimes 1 \right),
\end{eqnarray}
with $P^{R} + P^{L} = 1\otimes 1$. So the Dirac spinor is a superposition of these spinors
\begin{eqnarray}
\Psi^{Cl_{1,3}^{+} \otimes Cl_{1,3}^{+} } & = &\Psi_{R}^{Cl_{1,3}^{+} \otimes Cl_{1,3}^{+} }+\Psi_{L}^{Cl_{1,3}^{+} \otimes Cl_{1,3}^{+} },
\end{eqnarray}
wherein
\begin{eqnarray}
P^{R} \Psi^{Cl_{1,3}^{+} \otimes Cl_{1,3}^{+} } & = & \Psi_{R}^{Cl_{1,3}^{+} \otimes Cl_{1,3}^{+} } \nonumber \\
&=&[\alpha_{1} (\gamma_{3}\gamma_{0} \otimes \gamma_{3}\gamma_{0})
+ \alpha_{2}(\gamma_{3}\gamma_{0} \otimes \gamma_{1}\gamma_{0}\gamma_{2}\gamma_{0}) \nonumber \\
& & +\alpha_{3} (\gamma_{3}\gamma_{0} \otimes \gamma_{1}\gamma_{0})
+\alpha_{4} (\gamma_{3}\gamma_{0} \otimes \gamma_{2}\gamma_{0} \gamma_{3}\gamma_{0})]P^{\otimes 2}
\end{eqnarray}
and
\begin{eqnarray}
P^{L} \Psi^{Cl_{1,3}^{+} \otimes Cl_{1,3}^{+} } & = & \Psi_{L}^{Cl_{1,3}^{+} \otimes Cl_{1,3}^{+} } \nonumber \\
 &=&[\alpha_{5} (\gamma_{1}\gamma_{0} \otimes \gamma_{3}\gamma_{0})
 + \alpha_{6}(\gamma_{1}\gamma_{0} \otimes \gamma_{1}\gamma_{0}\gamma_{2}\gamma_{0}) \nonumber \\
& & + \alpha_{7} (\gamma_{1}\gamma_{0} \otimes \gamma_{1}\gamma_{0})
+\alpha_{8} (\gamma_{1}\gamma_{0}\otimes \gamma_{2}\gamma_{0} \gamma_{3}\gamma_{0})]P^{\otimes 2},
\end{eqnarray}
using that $\gamma_{3} \gamma_{0} P = P$. It is easy to see that $\Psi_{R}^{Cl_{1,3}^{+}}$ and $\Psi_{L}^{Cl_{1,3}^{+}}$ are not entangled states, and if state (\ref{bipartite}) is entangled the chiral projectors $P^{R}$ and $P^{L}$ always project such a state into separable states.

We will now analyze in the present context  the charge conjugation, parity and time reversal operations. In our formulation, they are unitary operators and can represent quantum logic gates. The parity operation, usually described in Dirac algebra by $\Phi_{P}=e^{i\phi}\gamma_{0}$, in our formulation is given by
\begin{eqnarray}
\Phi_{P}^{Cl_{1,3}^{+}\otimes Cl_{1,3}^{+}} & = & \left[ 1 \otimes (\cos(\phi)+\iota \sin(\phi)) \right]
\left[ \gamma_{3}\gamma_{0} \otimes 1 \right] ,
\end{eqnarray}
so the action of $\Phi_{P}^{Cl_{1,3}^{+}\otimes Cl_{1,3}^{+}}$ on $\Psi^{Cl_{1,3}^{+} \otimes Cl_{1,3}^{+} }$  is given by:

\begin{eqnarray}
\Phi_{P}^{Cl_{1,3}^{+}\otimes Cl_{1,3}^{+}} \Psi^{Cl_{1,3}^{+} \otimes Cl_{1,3}^{+} } &=&[1 \otimes (\cos(\phi)+\iota \sin(\phi))] \nonumber \\
& & \times \left[ \alpha_{1} (\gamma_{3}\gamma_{0} \otimes \gamma_{3}\gamma_{0})
+ \alpha_{2} (\gamma_{3}\gamma_{0} \otimes \gamma_{1}\gamma_{0}\gamma_{2}\gamma_{0}) \right. \nonumber \\
& & + \alpha_{3} (\gamma_{3}\gamma_{0} \otimes \gamma_{1}\gamma_{0})
+ \alpha_{4} (\gamma_{3}\gamma_{0} \otimes  \gamma_{2}\gamma_{0} \gamma_{3}\gamma_{0}) \nonumber \\
& & - \alpha_{5} (\gamma_{1}\gamma_{0} \otimes \gamma_{3}\gamma_{0})
- \alpha_{6} (\gamma_{1}\gamma_{0} \otimes \gamma_{1}\gamma_{0}\gamma_{2}\gamma_{0}) \nonumber \\
& & \left. - \alpha_{7} (\gamma_{1}\gamma_{0} \otimes  \gamma_{1}\gamma_{0})
- \alpha_{8} (\gamma_{1}\gamma_{0} \otimes \gamma_{2}\gamma_{0}\gamma_{3}\gamma_{0}) \right] P^{\otimes 2}, \nonumber \\
\end{eqnarray}
and we have that
\begin{eqnarray}
\widetilde{\Phi_{P}}^{Cl_{1,3}^{+}\otimes Cl_{1,3}^{+}}=[1 \otimes (\cos(\phi)-\iota \sin(\phi))](\gamma_{3}\gamma_{0} \otimes 1).
\end{eqnarray}

The charge conjugation $C=i \gamma_{2}\gamma_{0}$ can be written as:
\begin{eqnarray}
C^{Cl_{1,3}^{+}\otimes Cl_{1,3}^{+}} & = & (- \iota \otimes 1) (\gamma_{1}\gamma_{0} \otimes \gamma_{2}\gamma_{0})  \nonumber \\
& = & \gamma_{3}\gamma_{0} \gamma_{2}\gamma_{0} \otimes \gamma_{2}\gamma_{0} ,
\end{eqnarray}
with
$\widetilde{C}^{Cl_{1,3}^{+}\otimes Cl_{1,3}^{+}} = - C^{Cl_{1,3}^{+}\otimes Cl_{1,3}^{+}} $.
The action of the charge conjugation operator on the base states is given by:
\begin{eqnarray}
C^{Cl_{1,3}^{+}\otimes Cl_{1,3}^{+}}(\gamma_{3}\gamma_{0}\otimes \gamma_{3}\gamma_{0})P^{\otimes 2}
&=& - (\gamma_{2}\gamma_{0}\otimes \gamma_{2}\gamma_{0}\gamma_{3}\gamma_{0})P^{\otimes 2} \nonumber \\
& = & (\gamma_{1}\gamma_{0} \otimes \gamma_{1}\gamma_{0})P^{\otimes 2} ,
\end{eqnarray}
\begin{eqnarray}
C^{Cl_{1,3}^{+}\otimes Cl_{1,3}^{+}}(\gamma_{1}\gamma_{0}\otimes \gamma_{1}\gamma_{0})P^{\otimes 2}
& = & (\gamma_{1}\gamma_{0}\gamma_{2}\gamma_{0}\gamma_{3}\gamma_{0} \otimes \gamma_{1}\gamma_{0} \gamma_{2}\gamma_{0})P^{\otimes 2}. \nonumber \\
& = & - (\gamma_{3}\gamma_{0}\otimes \gamma_{3}\gamma_{0})P^{\otimes 2}.
\end{eqnarray}
\begin{eqnarray}
C^{Cl_{1,3}^{+}\otimes Cl_{1,3}^{+}}(\gamma_{3}\gamma_{0} \otimes \gamma_{1}\gamma_{0})P^{\otimes 2}
& = & (\gamma_{2}\gamma_{0} \otimes \gamma_{1}\gamma_{0} \gamma_{2}\gamma_{0})P^{\otimes 2} \nonumber \\
& = & -(\gamma_{1}\gamma_{0} \otimes \gamma_{3}\gamma_{0})P^{\otimes 2}
\end{eqnarray}
\begin{eqnarray}
C^{Cl_{1,3}^{+}\otimes Cl_{1,3}^{+}}(\gamma_{1}\gamma_{0}\otimes \gamma_{3}\gamma_{0})P^{\otimes 2}
& = & - (\gamma_{1}\gamma_{0}\gamma_{2}\gamma_{0}\gamma_{3}\gamma_{0}\otimes \gamma_{2}\gamma_{0}\gamma_{3}\gamma_{0})P^{\otimes 2} \nonumber \\
& = & (\gamma_{3}\gamma_{0} \otimes \gamma_{1}\gamma_{0})P^{\otimes 2}
\end{eqnarray}
using equivalent representations for the algebraic state. We can verify that they conduct entangled Bell states in themselves:
\begin{eqnarray}
C^{Cl_{1,3}^{+}\otimes Cl_{1,3}^{+}}\frac{1}{\sqrt{2}}[\gamma_{3}\gamma_{0}\otimes \gamma_{3}\gamma_{0} + \gamma_{1}\gamma_{0}\otimes \gamma_{1}\gamma_{0}]P^{\otimes 2}
& = & \frac{1}{\sqrt{2}} [\gamma_{1}\gamma_{0}\otimes \gamma_{1}\gamma_{0}-\gamma_{3}\gamma_{0}\otimes \gamma_{3}\gamma_{0}]P^{\otimes 2} \nonumber \\
C^{Cl_{1,3}^{+}\otimes Cl_{1,3}^{+}}\frac{1}{\sqrt{2}}[\gamma_{3}\gamma_{0}\otimes \gamma_{1}\gamma_{0}+\gamma_{1}\gamma_{0}\otimes \gamma_{3}\gamma_{0}]P^{\otimes 2}
& = & \frac{1}{\sqrt{2}}[\gamma_{3}\gamma_{0}\otimes \gamma_{1}\gamma_{0}-\gamma_{1}\gamma_{0}\otimes \gamma_{3}\gamma_{0}]P^{\otimes 2}. \nonumber \\
\end{eqnarray}

The time reversal $T=-i \gamma_{1}\gamma_{3}$ is given by
\begin{eqnarray}
T^{Cl_{1,3}^{+}\otimes Cl_{1,3}^{+}} =1 \otimes \gamma_{2}\gamma_{0}=\widetilde{T}^{Cl_{1,3}^{+}\otimes Cl_{1,3}^{+}}.
\end{eqnarray}
The effect of time reversal on base states is given by:
\begin{eqnarray}
T^{Cl_{1,3}^{+}\otimes Cl_{1,3}^{+}}(\gamma_{3}\gamma_{0}\otimes \gamma_{3}\gamma_{0})P^{\otimes 2}
& = & (1 \otimes \iota) (\gamma_{3}\gamma_{0}\otimes \gamma_{1}\gamma_{0})P^{\otimes 2} \nonumber \\
T^{Cl_{1,3}^{+}\otimes Cl_{1,3}^{+}}(\gamma_{1}\gamma_{0}\otimes \gamma_{1}\gamma_{0})P^{\otimes 2}
& = & -(1 \otimes \iota) (\gamma_{1}\gamma_{0} \otimes \gamma_{3}\gamma_{0})P^{\otimes 2} \nonumber \\
T^{Cl_{1,3}^{+}\otimes Cl_{1,3}^{+}}(\gamma_{3}\gamma_{0} \otimes \gamma_{1}\gamma_{0})P^{\otimes 2}
& = & -(1 \otimes \iota)  (\gamma_{3}\gamma_{0}\otimes \gamma_{3}\gamma_{0})P^{\otimes 2} \nonumber \\
T^{Cl_{1,3}^{+}\otimes Cl_{1,3}^{+}}(\gamma_{1}\gamma_{0}\otimes \gamma_{3}\gamma_{0})P^{\otimes 2}
& = & (1 \otimes \iota) (\gamma_{1}\gamma_{0} \otimes \gamma_{1}\gamma_{0})P^{\otimes 2} .
\end{eqnarray}
Entangled states are also preserved by the action of time reversal
\begin{eqnarray}
T^{Cl_{1,3}^{+}\otimes Cl_{1,3}^{+}}\frac{1}{\sqrt{2}}[\gamma_{3}\gamma_{0}\otimes \gamma_{3}\gamma_{0}+\gamma_{1}\gamma_{0}\otimes \gamma_{1}\gamma_{0}]P^{\otimes 2}
& = & \frac{ (1 \otimes \iota) }{\sqrt{2}}   \left[ \gamma_{3}\gamma_{0}\otimes \gamma_{1}\gamma_{0} \right.\nonumber \\
& & \left. - \gamma_{1}\gamma_{0}\otimes \gamma_{3}\gamma_{0}  \right] P^{\otimes 2}   \nonumber \\
T^{Cl_{1,3}^{+}\otimes Cl_{1,3}^{+}}\frac{1}{\sqrt{2}}[\gamma_{3}\gamma_{0}\otimes \gamma_{1}\gamma_{0}+\gamma_{1}\gamma_{0}\otimes \gamma_{3}\gamma_{0}]P^{\otimes 2}
& = & \frac{1}{\sqrt{2}} (1 \otimes \iota)   \left[ \gamma_{1}\gamma_{0}\otimes \gamma_{1}\gamma_{0} \right. \nonumber \\
& & \left. - \gamma_{3}\gamma_{0}\otimes \gamma_{3}\gamma_{0} \right] P^{\otimes 2}. \nonumber \\
\end{eqnarray}

We explore here the fact that Clifford algebras appear naturally in relativistic quantum mechanics through Dirac matrices $\gamma $. We believe that these results in terms of Clifford algebras and algebraic spinors without the use of a particular representation can  allow a better analysis of problems related to quantum information.


\section{Supersymmetry \label{sec5}}

Clifford algebras are essential tools in the construction of supersymmetric theories. Supersymmetry connects bosons and fermions and exhibits characteristics of major importance for theoretical physics, such as, for example, the solution to the hierarchy problem. Despite their theoretical relevance, supersymmetric partners were not observed. However, supersymmetry techniques have been used successfully in condensed matter physics, atomic physics and nuclear physics, for example. Clifford's algebras allow the construction of supersymmetric extensions of Poincar\'e's algebra, that are Lie superalgebras. \

In this section, motivated by the development carried out in the reference \cite{Lasenby}, in which an association between $ Cl_{1,3}$ algebra and supersymmetry was presented, we propose a new construction showing its relationship with quantum computing.
Thus, consider the $Q_{\alpha} $ supersymmetry generators in Lie-Poincar\'e superalgebra given by\cite{Lasenby}:
\begin{eqnarray}
Q_{\alpha}=-i\left(\frac{\partial}{\partial \theta^{\alpha}}-i\sigma_{\alpha \alpha '}^{\mu}\overline{\theta}^{\alpha'}\partial_{\mu}\right),
\end{eqnarray}
where $\mu$ is a spatial index, $\theta^{\alpha}$ and $\overline{\theta}^{\alpha}$ are Grassmann variables. It can be shown that by defining $\{\Gamma, \Gamma'\}\equiv \frac{1}{2} ( \Gamma \widetilde{\Gamma}'+ \widetilde{\Gamma} \Gamma')$ with
\begin{eqnarray}
\theta_{1}&=&1+\gamma_{3}\gamma_{0} \nonumber \\
\overline{\theta_{1}}&=&1-\gamma_{3}\gamma_{0} \nonumber \\
\theta_{2}&=&\gamma_{1}\gamma_{0}+\iota \gamma_{2}\gamma_{0} \nonumber \\
\overline{\theta_{2}}&=&-\gamma_{1}\gamma_{0}+\iota \gamma_{2}\gamma_{0},
\end{eqnarray}
we have
\begin{eqnarray}
\{\theta_{\alpha}, \theta_{\beta}\}= \{\overline{\theta}_{\alpha}, \overline{\theta}_{\beta}\}; \  \{\theta_{\alpha}, \overline{\theta}_{\beta}\}=2 \delta_{\alpha, \beta}, \ \ \ \textrm{with } \alpha,\beta = 1,2.
\end{eqnarray}
It is interesting to verify the action of these objects on the base elements $\gamma_{3}\gamma_{0}P, \gamma_{1}\gamma_{0}P$:
\begin{eqnarray}
\frac{1}{2}\theta_{1}\gamma_{3}\gamma_{0}P&=&\gamma_{3}\gamma_{0}P, \nonumber \\
\frac{1}{2}\overline{\theta}_{1}\gamma_{3}\gamma_{0}P&=&0 \cdot  \gamma_{3}\gamma_{0}P, \nonumber \\
\frac{1}{2}\theta_{2}\gamma_{3}\gamma_{0}P&=&0 \cdot  \gamma_{1}\gamma_{0}P, \nonumber \\
\frac{1}{2}\overline{\theta}_{2}\gamma_{3}\gamma_{0}P&=&- \gamma_{1}\gamma_{0}P.
\end{eqnarray}
and
\begin{eqnarray}
\frac{1}{2}\theta_{1}\gamma_{1}\gamma_{0}P&=&0 \cdot  \gamma_{1}\gamma_{0}P, \nonumber \\
\frac{1}{2}\overline{\theta}_{1}\gamma_{1}\gamma_{0}P&=& \gamma_{1}\gamma_{0}P, \nonumber \\
\frac{1}{2}\theta_{2}\gamma_{1}\gamma_{0}P&=& \gamma_{3}\gamma_{0}P, \nonumber \\
\frac{1}{2}\overline{\theta}_{2}\gamma_{1}\gamma_{0}P&=&0 \cdot  \gamma_{1}\gamma_{0}P.
\end{eqnarray}

A matrix representation for these operators using the algebraic base $\gamma_{3}\gamma_{0}P$ and $\gamma_{3}\gamma_{0}P$ is given by
\begin{eqnarray}
\theta _{1}=\left(
\begin{array}{cc}
2 & 0 \\
0 & 0%
\end{array}%
\right), \ \
\overline{\theta}_{1}=\left(
\begin{array}{cc}
0 & 0 \\
0 & 2%
\end{array}%
\right)
\end{eqnarray}
and
\begin{eqnarray}
\theta _{2}=\left(
\begin{array}{cc}
0 & 0 \\
2 & 0%
\end{array}%
\right), \ \
\overline{\theta}_{2}=\left(
\begin{array}{cc}
0 & -2 \\
0 & 0%
\end{array}%
\right).
\end{eqnarray}

Still in a supersymmetric context, using an octonionic formulation of the M-theory \cite{Anast}, the charge conjugation matrix $ \widehat{C}$ is given by:

\begin{eqnarray}
\widehat{C}=\left(
\begin{array}{cccc}
0 & 0 & 1 & 0 \\
0 & 0 & 0 & 1 \\
-1 & 0 & 0 & 0 \\
0 & -1 & 0 & 0%
\end{array}%
\right).
\end{eqnarray}
In our formalism we have
\begin{eqnarray}
\widehat{C}^{Cl_{1,3}^{+}\otimes Cl_{1,3}^{+}}=-(1 \otimes \iota) (\gamma_{2}\gamma_{0}  \otimes 1)=-\widetilde{\widehat{C}}^{Cl_{1,3}^{+}\otimes Cl_{1,3}^{+}}.
\end{eqnarray}
Its action on Bell states is given by
\begin{eqnarray}
\widehat{C}^{Cl_{1,3}^{+}\otimes Cl_{1,3}^{+}}\frac{1}{\sqrt{2}}[\gamma_{3}\gamma_{0}\otimes \gamma_{3}\gamma_{0}+\gamma_{1}\gamma_{0}\otimes \gamma_{1}\gamma_{0}]P^{\otimes 2}
& = & \frac{1}{\sqrt{2}}[\gamma_{1}\gamma_{0}\otimes \gamma_{3}\gamma_{0}-\gamma_{3}\gamma_{0}\otimes \gamma_{1}\gamma_{0}]P^{\otimes 2} \nonumber \\
\widehat{C}^{Cl_{1,3}^{+}\otimes Cl_{1,3}^{+}}\frac{1}{\sqrt{2}}[\gamma_{3}\gamma_{0}\otimes \gamma_{1}\gamma_{0}+\gamma_{1}\gamma_{0}\otimes \gamma_{3}\gamma_{0}]P^{\otimes 2}
& = & \frac{1}{\sqrt{2}} [\gamma_{1}\gamma_{0}\otimes \gamma_{1}\gamma_{0}-\gamma_{3}\gamma_{0}\otimes \gamma_{3}\gamma_{0}]P^{\otimes 2}. \nonumber \\
\end{eqnarray}
 Therefore these operators preserve entanglement. This prescription is valid for a larger number of dimensions, extending, as mentioned above, the number of factors in the tensor product of algebras.

\section{M-superalgebra \label{sec6}}

In M-theory, the most general supersymmetry algebra in $D=11$ Minkowskian spacetime is given by \cite{Anast,Toppan,Std}
\begin{eqnarray}
\{Q_{r},Q_{s}\}=(\widehat{C} \Gamma_{\mu})_{r,s}P^{\mu}+(\widehat{C} \Gamma_{[ \mu \nu ]})_{r,s}Z^{\mu \nu}+(\widehat{C} \Gamma_{[ \mu_{1}...\mu_{5} ]})_{r,s}Z^{\mu_{1}...\mu_{5}}, \label{eq-82}
\end{eqnarray}
where $Q$ is a pseudo-Majorana spinor that has $32$ components. Here we must have in the left hand side a symmetric matrix with 528 components; so, in the right-hand side, elements of rank $1$, $2$ and $5$ of Clifford algebra must constitute a basis for symmetric matrices \cite{Anast}. In fact, we have \cite{Toppan}
\begin{displaymath}
{11 \choose 1}+{11 \choose 2}+{11 \choose 5}=528.
\end{displaymath}
$\widehat{C}$ in equation (\ref{eq-82}) is charge conjugation matrix, $P^{\mu}$ is Poincar\'e generator of translation, $Z^{\mu \nu}$ is tensorial central charge of rank $2$, $Z^{\mu_{1}...\mu_{5}}$ is the tensorial central charge of rank $5$ and
\begin{eqnarray}
\Gamma_{[\mu_{1}...\mu_{l}]}=\frac{1}{l!}\sum_{\sigma \in G_{l}} \epsilon (\sigma) \Gamma_{\mu_{\sigma(1)}}...\Gamma_{\mu_{\sigma(l)}},
\end{eqnarray}
where $ \epsilon (\sigma)$ is the signature of permutation $\sigma$, $G_{p}$ is the symmetric group and, in the present section, $\Gamma_{ \mu_{\sigma_{1}} }$ is used to denote Dirac matrices (see below).
The charge conjugation must satisfy the relations \cite{Traubenberg}
\begin{eqnarray}
\widehat{C} \Gamma^{\mu}\widehat{C}^{T}=\xi (\Gamma^{\mu})^{T} \ \ \widehat{C}^{T}=\alpha \widehat{C},
\end{eqnarray}
where $\xi=(-1)^{p-1}$, $\widehat{C}^{T}$ is the transposed matrix $\widehat{C}$ and $\alpha=(-1)^{p(p-1)/2}$, with $(p,q)$ is the signature of the algebra. It was obtained in the reference \cite{Toppan} representations of $Cl_{p,q}$ through
\begin{equation}
\Gamma_{i}=\left(
\begin{array}{cc}
0 & \gamma_{i} \\
\gamma_{i} & 0%
\end{array}%
\right) ,\ \ \ \ \ \ \tau_{2}=\left(
\begin{array}{cc}
0 & 1_{d} \\
-1_{d} & 0%
\end{array}%
\right) ,\ \ \ \ \ \ \tau_{3}=\left(
\begin{array}{cc}
1_{d} & 0 \\
0 & -1_{d}%
\end{array}%
\right),
\end{equation}%
if $(p,q)\mapsto (p+1,q+1)$, defining $d$ as the dimension of $\gamma$,  or through
\begin{equation}
\Gamma_{i}=\left(
\begin{array}{cc}
0 & \gamma_{i} \\
-\gamma_{i} & 0%
\end{array}%
\right) ,\ \ \ \ \ \ \ \tau_{2}=\left(
\begin{array}{cc}
0 & 1_{d} \\
1_{d} & 0%
\end{array}%
\right) ,\ \ \ \ \ \ \tau_{3}=\left(
\begin{array}{cc}
1_{d} & 0 \\
0 & -1_{d}%
\end{array}%
\right).
\end{equation}%
if $(p,q)\mapsto (q+2,p)$, i.e. these $\Gamma_{i}$, $\tau_{2}$ and $\tau_{3}$ form a $2d$ dimensional representation of $Cl_{p,q}$. Indeed our construction can be applied to algebra $Cl_{0,7}$ and to obtain the algebras $Cl_{9,0}$ and $Cl_{10,1}$. We can build generators for $Cl_{10,1}$ through of generators of $Cl_{1,3}^{+}$ as

\begin{equation}
\Gamma_{\mu}=\left\{
\begin{array}{c}
\gamma _{1,0} \otimes \iota\gamma_{2,0} \otimes \gamma_{0,7}^{j} \\
\gamma _{1,0} \otimes \gamma _{1,0} \otimes 1 \otimes  1 \otimes 1 \\
\gamma _{1,0} \otimes \gamma _{3,0} \otimes 1 \otimes  1 \otimes 1 \\
\iota \gamma_{3,0} \otimes 1 \otimes 1 \otimes 1 \otimes 1\\
\gamma_{3,0} \otimes 1 \otimes 1 \otimes 1 \otimes 1
\end{array}
\right.
\end{equation}
where
\begin{equation}
\left\{
\begin{array}{c}
\gamma _{0,7}^{1}= \iota\gamma_{2,0} \otimes \gamma_{1,0} \otimes 1 \\
\gamma _{0,7}^{2}= \iota\gamma_{2,0} \otimes \gamma_{3,0} \otimes 1 \\
\gamma _{0,7}^{3}= 1 \otimes \iota\gamma_{2,0} \otimes \gamma_{1,0} \\
\gamma _{0,7}^{4}= 1 \otimes \iota\gamma_{2,0} \otimes \gamma_{3,0} \\
\gamma _{0,7}^{5}= \gamma_{1,0} \otimes 1 \otimes \iota\gamma_{2,0} \\
\gamma _{0,7}^{6}= \gamma_{3,0} \otimes 1 \otimes \iota\gamma_{2,0}\\
\gamma _{0,7}^{7}= \iota\gamma_{2,0} \otimes \iota\gamma_{2,0} \otimes \iota\gamma_{2,0}
\end{array}
\right.
\end{equation}
are generators of Clifford algebra $Cl_{0,7}$, where $\gamma_{i,0}\equiv \gamma_{i}\gamma_{0}$. The charge conjugation in our formulation is given by
\begin{eqnarray}
\widehat{C}^{Cl_{10,1}^{+}}=\gamma_{3,0} \otimes 1 \otimes 1 \otimes 1 \otimes 1.
\end{eqnarray}
If we think in terms of an algebraic spinor we can associate these spinors with a state of 5 qubits. Note that $Cl_{10,1}\subset Cl_{1,3}^{+ \otimes 5} \simeq Cl_{3,0}^{\otimes 5}$. If $S^{Cl_{10,1}}$ is the spinor representation for $Cl_{10,1}$, by using the results of references \cite{Traubenberg, Varad, Del}, we have
\begin{eqnarray}
End (S^{Cl_{10,1}}) &\simeq & S^{Cl_{10,1}} \otimes S^{Cl_{10,1}} \nonumber \\
 &\simeq & \bigwedge ^{(1)}(\mathbb{R}^{10,1}) \oplus \bigwedge ^{(2)}(\mathbb{R}^{10,1}) \oplus \bigwedge ^{(5)}(\mathbb{R}^{10,1})
\end{eqnarray}
where $End$ designates endomorphism. So we have,
\begin{eqnarray}
\Gamma_{[\mu_{1}\mu_{2}]}=\frac{1}{2!}\sum_{\sigma \in G_{p}} \epsilon (\sigma) \Gamma_{\mu_{\sigma(1)}}\Gamma_{\mu_{\sigma(2)}}
\end{eqnarray}
and
\begin{eqnarray}
\Gamma_{[\mu_{1}\mu_{2}\mu_{3}\mu_{4}\mu_{5}]}=\frac{1}{5!}\sum_{\sigma \in G_{p}} \epsilon (\sigma) \Gamma_{\mu_{\sigma(1)}}\Gamma_{\mu_{\sigma(2)}}\Gamma_{\mu_{\sigma(3)}}\Gamma_{\mu_{\sigma(4)}}\Gamma_{\mu_{\sigma(5)}}.
\end{eqnarray}
We can also build
\begin{eqnarray}
(S^{Cl_{3,0}} \otimes S^{Cl_{3,0}})^{\otimes 5}
 \simeq  (\bigwedge ^{(0)}(\mathbb{R}^{3,0}) \oplus \bigwedge ^{(2)}(\mathbb{R}^{3,0}))^{\otimes 5} \nonumber
\end{eqnarray}

We can show that $S^{Cl_{10,1}} \otimes S^{Cl_{10,1}} \subset (S^{Cl_{3,0}} \otimes S^{Cl_{3,0}})^{\otimes 5}$ for some spinor space $S^{Cl_{3,0}^{\otimes 5}}$. For this we note that we have $S^{Cl_{10,1}} \simeq I_{10,1}$, a minimal left ideal of $Cl_{10,1}$ and similarly $S^{Cl_{3,0}^{\otimes 5}} \simeq I_{Cl_{3,0}^{\otimes 5}}$, a minimal left ideal of $Cl_{3,0}^{\otimes 5}$. Every ideal of an algebra $A$ may be written as $AE$, where $E$ is a primitive idempotent. Therefore, we have $I_{10,1}=Cl_{10,1}E_{10,1}$ and $I_{Cl_{3,0}^{\otimes 5}}=Cl_{3,0}^{\otimes 5}E_{Cl_{3,0}^{\otimes 5}}$ Since both algebras have the same dimension related to irreducible representation, each primitive idempotent of $Cl_{10,1}$ corresponds to a primitive idempotent of $Cl_{3,0}^{\otimes 5}$. Remembering that primitive idempotents always admit a diagonal representation and $Cl_{10,1}\subset Cl_{1,3}^{+ \otimes 5}$ the proof is finished.
\
The M-superalgebra also admits an octonionic representation \cite{Anast,Toppan,Top2}. Consider octonions $o_{i}$ satisfying
\begin{eqnarray}
o_{i}\circ o_{j}=- \delta_{i,j} + \overline{C}_{i,j,k}, \ \ i,j,k=1,...,7
\end{eqnarray}
where $\overline{C}_{i,j,k}$ are totally antisymmetric octonionic structure constants \cite{Toppan}. We set $o_{i}=\gamma_{0,7}^{i} \in Cl_{0,7}$. A general octonion is given by
\begin{eqnarray}
o=\lambda_{0}o_{0}+\lambda_{1}o_{1}+\lambda_{2}o_{2}+\lambda_{3}o_{3}+\lambda_{4}o_{4}+\lambda_{5}o_{5}+\lambda_{6}o_{6}+\lambda_{7}o_{7},
\end{eqnarray}
with $\lambda_{k} \in \mathbb{R}; \ \ k=0,...,7$. The conjugate $o*$ of octonion is given by
\begin{eqnarray}
o=\lambda_{0}o_{0}-\lambda_{1}o_{1}-\lambda_{2}o_{2}-\lambda_{3}o_{3}-\lambda_{4}o_{4}-\lambda_{5}o_{5}-\lambda_{6}o_{6}-\lambda_{7}o_{7}.
\end{eqnarray}
The norm is defined as
\begin{eqnarray}
\|o\| & = & \sqrt{oo*} = \sqrt{\lambda_{0}^{2}+\lambda_{1}^{2}+\lambda_{2}^{2}+\lambda_{3}^{2}+\lambda_{4}^{2}+\lambda_{5}^{2}+\lambda_{6}^{2} +\lambda_{7}^{2}  } \nonumber
\end{eqnarray}
In our formulation an octonionic representation of $Cl_{10,1}$ can be build as

\begin{eqnarray}
\Gamma_{10,1}^{o_{i}}=\left(
\begin{array}{cccc}
0 & 0 & 0 & o_{i} \\
0 & 0 & -o_{i} & 0 \\
0 & o_{i} & 0 & 0 \\
-o_{i} & 0 & 0 & 0%
\end{array}%
\right) \ \ \ \textrm{for } i=1,..7,
\end{eqnarray}

\begin{eqnarray}
\Gamma_{10,1}^{o,8}=\left(
\begin{array}{cccc}
0 & 0 & 0 & 1 \\
0 & 0 & 1 & 0 \\
0 & 1 & 0 & 0 \\
1 & 0 & 0 & 0%
\end{array}%
\right),
\qquad
\Gamma_{10,1}^{o,9}=\left(
\begin{array}{cccc}
0 & 0 & 1 & 0 \\
0 & 0 & 0 & -1 \\
1 & 0 & 0 & 0 \\
0 & -1 & 0 & 0%
\end{array}%
\right),
\end{eqnarray}

\begin{eqnarray}
\Gamma_{10,1}^{o,10}=\left(
\begin{array}{cccc}
1 & 0 & 0 & 0 \\
0 & 1 & 0 & 0 \\
0 & 0 & -1 & 0 \\
0 & 0 & 0 & -1%
\end{array}%
\right),
\qquad
\Gamma_{10,1}^{o,0}=\left(
\begin{array}{cccc}
0 & 0 & 1 & 0 \\
0 & 0 & 0 & 1 \\
-1 & 0 & 0 & 0 \\
0 & -1 & 0 & 0%
\end{array}%
\right) ,
\end{eqnarray}
and we can write the Majorana spinor in $D=11$ as an $4 \times 4$ octonionic column vector
\begin{eqnarray}\label{smajorana}
\zeta_{10,1}^{o,0}=\left(
\begin{array}{cccc}
\zeta_{1} \\
\zeta_{2} \\
\zeta_{3} \\
\zeta_{4} %
\end{array}%
\right),
\end{eqnarray}
with $\zeta_{k} \in O, \ \ k=1,..,4$. We define the octonionic qubit as

\begin{eqnarray}
|\vartheta \rangle_{o}=\left(
\begin{array}{cccc}
\vartheta_{1} \\
\vartheta_{2}%
\end{array}%
\right),
\end{eqnarray}
where $\vartheta_{k} \in O, \ \ k=1,2$ and $\| \vartheta_{1}\|^{2}+\| \vartheta_{2}\|^{2}=1$. Consequently it is possible to identify the spinor (\ref{smajorana}) as a state of two octonionic qubits. Obtaining a separability criterion for octonionic qubits is not so simple because octonions are noncommutative and nonassociative. However they constitute an algebra without zero divisor so that we can verify, for example, that the state

\begin{eqnarray}
\zeta_{10,1}^{o,0}=\left(
\begin{array}{cccc}
\zeta_{a} \\
0 \\
0 \\
\zeta_{b} %
\end{array}%
\right),
\end{eqnarray}
is entangled.

\section{Conclusions\label{sec7}}
In this paper, we have derived a formulation based on tensor product of Clifford algebras and minimal left ideals. In this context, we have show how to build quantum logic gates and general states of a qubits.
Furthermore, we have shown as concepts from particle physics such as chirality, charge conjugation, parity and time reversal are easily introduced with our formulation and can be expressed in connection to quantum information theory.  Such a relationship is possible by noting that Dirac algebra is a Clifford algebra. Subsequently, we have shown how to obtain Grassmann variable in a supersymmetry context as well as the charge conjugation matrix in an algebraic formulation of M-theory.
We note that the derived operators are Hermitian or unitary operators corresponding to  quantum observables and logic gates, respectively. All states and operators are described in terms of the algebra generators. The M-algebra were derived from Clifford algebra $Cl_{1,3}^{+}\simeq Cl_{3,0}$ and its relationship between qubits and octonions was analyzed. It is our view that these results may be useful in future developments of quantum computing related to simulations of relativistic systems, high energy physics, and M-theory.


\end{document}